\documentclass[review,12pt]{elsarticle}
\usepackage[T1]{fontenc}
\usepackage{lmodern}
\usepackage{amsmath,amssymb,amsthm}
\usepackage{graphicx,booktabs}
\usepackage{float,enumitem}
\floatstyle{ruled}
\newfloat{algorithm}{tbp}{loa}
\floatname{algorithm}{Algorithm}
\usepackage[hidelinks]{hyperref}
\newtheorem{theorem}{Theorem}[section]
\newtheorem{lemma}[theorem]{Lemma}
\newtheorem{corollary}[theorem]{Corollary}

\newtheorem{proofclaim}{Claim}
\theoremstyle{definition}

\newtheorem{example}[theorem]{Example}
\newtheorem{assumption}{Assumption}
\theoremstyle{remark}

\newcommand{\TT}{\mathcal T}
\newcommand{\MT}{M^T}
\newcommand{\HT}{H^T}
\newcommand{\NT}{\mathcal{N}^T}
\newcommand{\PP}{\mathcal P^T}
\newcommand{\gx}{\widehat x}
\newcommand{\OGFPT}{\mathrm{OGFP}^T}
\journal{Theoretical Computer Science}
\hypersetup{pdftitle={Generalized Flows over Time: Complexity, Structure and Algorithm},pdfauthor={Siyuan Chen, Shengminjie Chen, Suixiang Gao, Zheyu Jiang, Dun Ma, Wenguo Yang}}

\begin{document}
	\begin{frontmatter}
		\title{Generalized Flows over Time: Complexity, Structure and Algorithm\tnoteref{conference}}
		\tnotetext[conference]{An earlier version of this paper appeared at COCOON 2026~\cite{chen2027structure}.}
		\author[aff1]{Siyuan~Chen}
		\ead{chensiyuan221@mails.ucas.ac.cn}
		
		\author[aff2,aff3]{Shengminjie~Chen}
		\ead{csmj@ict.ac.cn}
		
		\author[aff1,aff4]{Suixiang~Gao\corref{cor1}}
		\ead{sxgao@ucas.ac.cn}
		
		\author[aff1]{Zheyu~Jiang}
		\ead{jiangzheyu24@mails.ucas.ac.cn}
		
		\author[aff5]{Dun~Ma}
		\ead{madun21@mails.ucas.ac.cn}
		
		\author[aff1]{Wenguo~Yang}
		\ead{yangwg@ucas.ac.cn}
		
		\affiliation[aff1]{
			organization={School of Mathematical Sciences,
				University of Chinese Academy of Sciences},
			city={Beijing},
			postcode={100049},
			country={China}
		}
		
		\affiliation[aff2]{
			organization={State Key Lab of Processors,
				Institute of Computing Technology,
				Chinese Academy of Sciences},
			city={Beijing},
			postcode={100190},
			country={China}
		}
		
		\affiliation[aff3]{
			organization={School of Computer Science and Technology,
				University of Chinese Academy of Sciences},
			city={Beijing},
			postcode={100049},
			country={China}
		}
		
		\affiliation[aff4]{
			organization={Zhongguancun Laboratory},
			city={Beijing},
			postcode={100000},
			country={China}
		}
		
		\affiliation[aff5]{
			organization={School of Advanced Interdisciplinary Sciences,
				University of Chinese Academy of Sciences},
			city={Beijing},
			postcode={101408},
			country={China}
		}
		
		\cortext[cor1]{Corresponding author.}

		\begin{abstract}
			Generalized flows over time constitute a general network flow model that incorporates both transit times and flow losses or gains. We study their complexity, structure, and approximation. We prove that the generalized maximum flow over time problem remains NP-hard under standard binary encoding, even on lossy two-terminal series-parallel networks. The maximum temporally repeated restriction is APX-hard with general gains and NP-hard on lossy networks. We then consider the minimum source supply needed to satisfy a fixed sink demand within a given horizon, and prove that an optimal solution always exists without intermediate storage. On lossy networks, such a solution can additionally use only simple source--sink paths. These structural properties yield a fully polynomial-time approximation scheme via time condensation, with analogous schemes for generalized maximum flows over time and quickest flows with a source-supply budget.
		\end{abstract}
	\end{frontmatter}
	\section{Introduction}\label{sec:introduction}
	Network flow is a fundamental class of problems in combinatorial optimization, with widespread applications across logistics, transportation, and telecommunications (see, e.g.,~\cite{ahuja1995network,williamson2019network}). Classical network flow models typically assume flow invariance along arcs and instantaneous transmission. However, in practical scenarios such as the long-distance pipeline transmission of oil or gas, the transportation process incurs not only physical losses or leakages but also transit times---characteristics that classical models fail to capture. To address these limitations, two significant generalizations have been established: \textit{generalized flows}, which characterize physical flow changes, and \textit{flows over time}, which account for the temporal dynamics of transmission. While research on these two extensions has historically proceeded in parallel, accurately modeling scenarios like pipeline transmission necessitates the integration of both frameworks.
	
	Gro\ss{} and Skutella~\cite{gross2011generalized} initiated the study of generalized flows over time. In their model, each arc $a$ is associated with a \textit{gain factor} $\gamma_a$ and a \textit{transit time} $\tau_a$. Specifically, $x$ units of flow entering arc $a$ at time $t$ will result in $\gamma_a x$ units exiting at time $t+\tau_a$. Within this framework, they investigated the \textit{generalized maximum flow over time problem} (GMFP), which seeks to maximize the total flow reaching the sink within a fixed time horizon $T$.
	
	Gro\ss{} and Skutella~\cite{gross2011generalized} established that the GMFP is NP-hard and even inapproximable. However, the gain factors in their reduction are implicitly encoded through the transit times, and their ordinary binary representations may have exponential length. Gro\ss{}~\cite[Section~7.1.1]{gross2014approximation} therefore left open whether the hardness persists under binary encoding. This leads to our first question:
	\begin{quote}
		\textit{Is the GMFP still NP-hard under standard binary encoding?}
	\end{quote}
	
	The authors~\cite{gross2011generalized} also designed a \textit{fully polynomial time approximation scheme} (FPTAS) for the GMFP on lossy networks with proportional losses. Gro\ss{}~\cite[Chapter~5]{gross2014approximation} subsequently developed time-condensation algorithms for general \textit{lossy networks}, where all gain factors are at most $1$. However, their model and algorithms allow flow to be stored at intermediate nodes. In many practical scenarios such as traffic networks, limited space or time resources make intermediate stops undesirable or even prohibited. This naturally raises a second question:
	\begin{quote}
		\textit{Does the GMFP always admit a storage-free optimal solution?}
	\end{quote}
	This work considers these two questions and gives affirmative answers to both.
	
	\paragraph{Our contributions}
	We first establish the NP-hardness of the GMFP under standard binary encoding via a reduction from the NP-complete problem \textsc{Exact Cover by 3-Sets} (X3C). This holds even on lossy \emph{two-terminal series-parallel} (TTSP) networks with unit capacities, resolving the open question raised by Gro\ss{}~\cite{gross2014approximation}.
	
	We then examine whether the GMFP admits a temporally repeated optimal solution, as such flows require no intermediate storage. A TTSP network with only three arcs shows that it does not. Furthermore, finding a generalized maximum temporally repeated flow is APX-hard with general gains and remains NP-hard on lossy TTSP networks.
	
	These negative results compel us to adopt a direct proof approach. We consider the \textit{optimal generalized flow over time problem} (OGFP), which minimizes the total source supply for a fixed sink demand and time horizon. We prove that it always admits an optimal solution without intermediate storage; taking the demand to be the optimal GMFP value gives the same conclusion for the GMFP. Our proof builds on the time-shift framework used by Fleischer and Skutella~\cite{fleischer2007quickest} to show that the minimum-cost flow over time problem admits a storage-free optimal solution. We extend this framework to accommodate gains and losses.
	
	Further, we prove that in lossy networks, an optimal OGFP solution can simultaneously avoid storage and use only simple source--sink paths. This property allows us to apply the time-condensation framework of Fleischer and Skutella~\cite{fleischer2007quickest} and Gro\ss{}~\cite{gross2014approximation} to obtain a storage-free FPTAS for the OGFP. Binary search gives the corresponding storage-free approximation for the GMFP and an FPTAS for the quickest-flow problem with a prescribed demand and source-supply budget. 
	
	\paragraph{Related work}
	Generalized flows were introduced by Kantorovich~\cite{kantorovich1960mathematical}; each arc is associated with a gain factor to model flow gain or loss. We refer the reader to Wayne~\cite{wayne1999generalized} and Lin~\cite{lin2001algorithms} for a comprehensive overview of generalized flow problems and related algorithms. The generalized maximum flow problem is the most classic and extensively studied subproblem. Goldberg, Plotkin, and Tardos~\cite{goldberg1991combinatorial} presented polynomial-time combinatorial algorithms, and Fleischer and Wayne~\cite{fleischer2002fast} designed fast approximation schemes for generalized flow problems. V\'egh~\cite{vegh2017strongly} presented the first strongly polynomial time algorithm for generalized flow maximization and subsequently, with Olver~\cite{olver2020simpler}, proposed a simpler and faster strongly polynomial time algorithm. Krumke and Zeck~\cite{krumke2013generalized} studied generalized max flow on series-parallel graphs.
	
	Parallel to the study of physical flow changes, Ford and Fulkerson~\cite{ford1958constructing} pioneered the investigation of flows over time (also known as dynamic flows), associating transit times with arcs. For thorough surveys on flows over time, we refer to Aronson~\cite{aronson1989survey} and Skutella~\cite{skutella2009introduction}. The time-expanded network~\cite{ford1962flows} is a fundamental tool for solving such problems, transforming a dynamic flow problem into an equivalent static flow problem on a larger network. However, since network size depends linearly on the time horizon $T$, algorithms based on it typically exhibit pseudopolynomial time complexity. To overcome this limitation, Fleischer and Skutella~\cite{fleischer2007quickest} introduced the time-condensation technique, which reduces the network size to polynomial scale at the cost of slightly relaxing the time horizon. Leveraging this technique, they designed FPTASs for numerous dynamic flow problems. Fleischer and Tardos~\cite{fleischer1998efficient} studied the relation between discrete- and continuous-time flows and developed efficient continuous-time algorithms. Klinz and Woeginger~\cite{klinz2004minimum} established the complexity of minimum cost dynamic flows on TTSP networks.
	
	The study of generalized flows over time was initiated by Gro\ss{} and Skutella~\cite{gross2011generalized} and developed further in Gro\ss{}'s thesis~\cite{gross2014approximation}. More recent work considers generalized maximum flows over time with intermediate storage~\cite{dhamala2024generalized} and generalized flows with excess storage on TTSP lossy networks~\cite{adhikari2025generalized}. These models allow stored excess flow to contribute to the objective. Chen et al.~\cite{chen2026earliest} study the existence and approximability of generalized earliest arrival flows, which impose the delivery objective simultaneously over time.
	
	Although existing results show that most single-commodity flow over time problems admit storage-free optimal solutions, the landscape differs in the multicommodity setting. In certain instances, allowing intermediate storage can indeed strictly improve the optimal solution, although this improvement is bounded, as discussed in detail by Gro\ss{} and Skutella~\cite{gross2015tight}. They also designed approximation schemes for maximum multicommodity flows over time without intermediate storage using sequence condensation~\cite{gross2012maximum}.

	\section{Preliminaries}\label{sec:preliminaries}
	\paragraph{Notation}
	Let $G=(V,A)$ be a directed graph, where $V$ is the node set with
	$n:=|V|$ and $A$ is the arc set with $m:=|A|$. We designate two distinct
	nodes $s,d\in V$ as the source and sink, respectively, and assume the source
	has no incoming arcs and the sink has no outgoing arcs. All nodes in
	$V\setminus\{s,d\}$ are called \emph{intermediate nodes}. For any node
	$i\in V$, let $\delta^+(i)$ and $\delta^-(i)$ denote the sets of arcs
	leaving and entering $i$, respectively. For any arc $a=(i,j)\in A$, we
	refer to $i$ as its \emph{tail} and $j$ as its \emph{head}. 
	
	By assigning attributes to the arcs of $G$, we obtain a generalized dynamic
	network $\mathcal{N}=(G,u,\tau,\gamma,s,d,T)$. Here, $u\in\mathbb R^m_{>0}$ assigns
	a finite positive capacity $u_a$ to each arc $a$;
	$\tau\in\mathbb Z^m_{\ge0}$ assigns an integer transit time $\tau_a$ to
	each arc, interpreted as the time required for flow to move from its tail
	to its head; and $\gamma\in\mathbb R^m_{>0}$ assigns a gain factor
	$\gamma_a$ to each arc, modeling the change in flow magnitude along the arc.
	Specifically, if $x$ units of flow enter arc $a$ at its tail, then
	$\gamma_a x$ units arrive at its head after $\tau_a$ time units.
	We call a network $\mathcal{N}$ \emph{lossy} if $\gamma_a\le1$ for all $a\in A$.
	Given an integer $T\ge1$, we study problems over the time horizon $[0,T)$.

	\paragraph{Walks, paths, and cycles}
	A \emph{walk} is a sequence of consecutively incident directed arcs.
	A \emph{path} visits no node twice, and a \emph{cycle} is a closed walk
	with no repeated node other than its first and last node.
	
	For a walk $P=(a_1,\ldots,a_k)$, define
	\begin{align*}
		\tau(P)&=\sum_{j=1}^k\tau_{a_j},&
		\gamma(P)&=\prod_{j=1}^k\gamma_{a_j},\\
		\tau(a_j\mid P)&=\sum_{r<j}\tau_{a_r},&
		\gamma(a_j\mid P)&=\prod_{r<j}\gamma_{a_r}.
	\end{align*}
	Empty sums and products equal $0$ and $1$, respectively. A path flow
	of source value $x_P$ uses $\gamma(a\mid P)x_P$ on arc $a$ and delivers
	$\gamma(P)x_P$ to the sink. A cycle is \emph{flow-generating},
	\emph{unit-gain}, or \emph{flow-absorbing} according as its gain is
	greater than, equal to, or less than $1$.
	
	\paragraph{Series-parallel graphs}
	In the sequel we will repeatedly use two-terminal series-parallel (TTSP)
	graphs. The simplest TTSP graph consists of a single arc $(s,d)$.
	These graphs are defined recursively. Let $G_1$ and $G_2$ be TTSP graphs
	with terminals $s_1,d_1$ and $s_2,d_2$, respectively. We obtain a new TTSP
	graph in either of the following ways:
	
	\emph{Series composition:} identify $d_1$ with $s_2$; the resulting graph
	has terminals $s_1,d_2$.
	
	\emph{Parallel composition:} identify $s_1$ with $s_2$ to form $s$, and
	identify $d_1$ with $d_2$ to form $d$; the resulting graph has terminals $s,d$.
	
\begin{figure}[t]
		\centering
		\includegraphics[width=.8\linewidth]{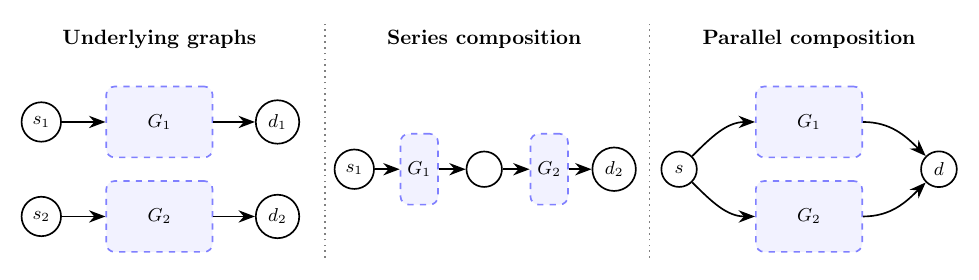}
		\caption{Series and parallel composition of two-terminal networks.}
		\label{pr:ttsp}
	\end{figure}

	Figure~\ref{pr:ttsp} illustrates both compositions. Valdes et
	al.~\cite{valdes1982recognition} presented a linear-time recognition
	algorithm for TTSP graphs.
	
	\paragraph{Generalized static flows}
	A generalized static flow is a function $x:A\to\mathbb R_{\ge0}$ satisfying
	\begin{align*}
		\sum_{a\in\delta^-(i)}\gamma_a x_a-\sum_{a\in\delta^+(i)}x_a&=0
		&&(i\in V\setminus\{s,d\}),\\
		0\le x_a&\le u_a &&(a\in A).
	\end{align*}
	Given a feasible flow $x$ in $G$, the \emph{residual graph} is
	$G_{x}=(V,A_{x}=A \cup \overleftarrow{A}).$
	For every forward arc $a\in A$, the residual capacity is
	$
	u^{x}_{a}:=u_{a}-x_{a}\geq 0,
	$
	which is the additional flow that can still be sent along $a$. 
	For every reverse arc $\overleftarrow{a}$, set its residual capacity to
	$
	u_{\overleftarrow{a}}:=\gamma_{a}\,x_{a}\ge 0,
	$
	representing the maximum amount of flow that can be pushed back from the head of $a$ to its tail.   It is easy to verify that if $y$ is a feasible flow in the residual graph, then the sum $x+y$ is again a feasible flow in $G$, with
	$
	(x+y)_a:=x_a+y_a-\gamma_{\overleftarrow{a}}\,y_{\overleftarrow{a}} \text{ for any } a\in A.
	$
	
	\paragraph{Generalized flows over time}
	A generalized flow over time is a Lebesgue-integrable function
	$f:A\times[0,T)\to\mathbb R_{\ge0}$. Here we work in a continuous-time
	model: $f_a(t)$ denotes the instantaneous inflow rate entering arc $a$
	at time $t$. For convenience, we adopt the convention $f_a(t)=0$ for
	$t<0$ or $t\ge T-\tau_a$. For an intermediate node $i$, define its
	inventory by
	\begin{equation}\label{pr:inventory}
		B_i(t)=\sum_{a\in\delta^-(i)}\int_0^t\gamma_a f_a(\theta-\tau_a)\,d\theta
		-\sum_{a\in\delta^+(i)}\int_0^t f_a(\theta)\,d\theta.
	\end{equation}
	A flow is \emph{feasible} if $0\le f_a(t)\le u_a$ almost everywhere,
	$B_i(t)\ge0$ for $0\le t\le T$, and $B_i(T)=0$ at every intermediate
	node. The inequality permits intermediate storage, while the final
	equality requires all intermediate inventories to be empty at completion.
	A flow is \emph{storage-free} if $B_i(t)=0$ for all intermediate $i$
	and all $t$. 
	
	Based on the aforementioned definition, we first restate the
	\emph{generalized maximum flow over time problem} (GMFP) with a given
	time horizon $T$, proposed in~\cite{gross2011generalized}:
	\[
	(\mathrm{GMFP})\qquad
	v^*(T)=\max_f\ \sum_{a\in\delta^-(d)}\int_0^T\gamma_a f_a(t)\,dt.
	\]
	Here and below, feasibility constraints are understood. 
	Although the GMFP is well-defined, flow gains and losses in generalized
	dynamic networks may cause a discrepancy between source outflow and sink
	inflow. For a fixed sink demand $b$ and time horizon $T$, cost efficiency
	motivates minimizing the total source supply. This leads to the
	\emph{optimal generalized flow over time problem} (OGFP):
	\begin{align*}
		(\mathrm{OGFP})\qquad S^*(b,T)
		&=\min_f\ \sum_{a\in\delta^+(s)}\int_0^T f_a(t)\,dt\\
		&\text{s.t.}\quad
		\sum_{a\in\delta^-(d)}\int_0^T\gamma_a f_a(t)\,dt=b.
	\end{align*}
	Set $S^*(b,T)=+\infty$ when demand $b$ is infeasible. The case $b=0$
	has the zero flow as an optimum. Taking $b=v^*(T)$ relates the two problems.
	
	\paragraph{Temporally repeated flows}
	Let $\mathcal P$ be the set of simple $s$--$d$ paths in $G$.
	A feasible collection of static path rates $x_P\ge0$ satisfies
	$\sum_{P:a \in P}\gamma(a\mid P)x_P\le u_a$ for each $a$.
	Its temporally repeated realization sends at rate $x_P$ along $P$
	during $[0,(T-\tau(P))_+)$, without waiting, where $(z)_+=\max\{z,0\}$.
	The \emph{generalized maximum temporally repeated flow problem} (GMTRFP)
	is the path-based optimization problem
	\[
	\max\left\{\sum_{P\in\mathcal P}(T-\tau(P))_+\gamma(P)x_P:
	\sum_{P\ni a}\gamma(a\mid P)x_P\le u_a\ (a\in A),\quad x_P\ge0\right\}.
	\]
	A solution is specified by its positive path rates. The static capacity
	inequalities ensure feasibility of its dynamic
	realization, since at every time only a subset of the path rates uses
	a given arc. 
	
	\paragraph{Time expansion and time condensation}
	Given a dynamic network $\mathcal{N}=(G,u,\tau,\gamma,s,d,T)$, we construct its
	time-expanded network $\NT$ as follows. For each $i\in V$ and
	$\theta\in\TT:=\{0,\ldots,T-1\}$, create a copy $i_\theta$.
	Its arc set consists of the \emph{movement arcs}
	\[
	\MT=\{a_\theta=(i_\theta,j_{\theta+\tau_a}):
	a=(i,j)\in A,\ 0\le\theta<T-\tau_a\}
	\]
	and the \emph{holdover arcs}
	\[
	\HT=\{(i_\theta,i_{\theta+1}):i\in V,\ 0\le\theta<T-1\}.
	\]
	Assign capacities and gains by
	\[
	u_{a'}=
	\begin{cases}
		u_a,&a'=a_\theta\in\MT,\\
		\infty,&a'\in\HT,
	\end{cases}
	\qquad
	\gamma_{a'}=
	\begin{cases}
		\gamma_a,&a'=a_\theta\in\MT,\\
		1,&a'\in\HT.
	\end{cases}
	\]
	Holdover arcs are included also at the source and sink. We therefore use
	$s_0$ and $d_{T-1}$ as the unique source and sink, respectively.
	Source holdovers carry supply reserved for later departures, while sink
	holdovers carry earlier arrivals. Let $\text{OGFP}^T$ be the counterpart of the OGFP on the time-expanded network $\mathcal{N}^T$.
	
\begin{figure}[t]
		\centering
		\includegraphics[width=\linewidth]{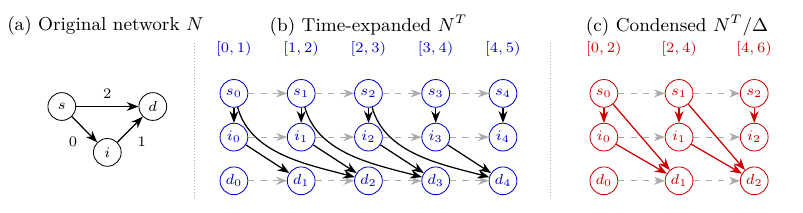}
		\caption{(a) The original network $\mathcal{N}$, with transit times on its arcs and
			unit capacities and gains; (b) its time-expanded network for $T=5$;
			(c) its condensed time-expanded network with $\Delta=2$, rounded horizon
			$6$, and movement capacities $2$. Dashed arcs are holdover arcs.}
		\label{pr:time-condensation}
	\end{figure}

	The size of the time-expanded network $\NT$ may grow exponentially
	compared with $\mathcal{N}$. To overcome this issue, a natural idea is to choose a
	larger time unit; for example, if time was previously measured in seconds,
	now measure it in minutes. Take a time unit $\Delta>0$ and round the arc
	transit times to $\tau'_a=\Delta\lceil\tau_a/\Delta\rceil$.
	Writing $\ell_a=\tau'_a/\Delta$ and $K=\lceil T/\Delta\rceil$, construct
	a $K$-layer network with node copies $i_k$, $0\le k<K$, movement arcs
	$(i_k,j_{k+\ell_a})$ whenever $k+\ell_a<K$, and holdover arcs between
	consecutive copies of every node.
	Movement arcs have capacity $\Delta u_a$ and gain $\gamma_a$;
	holdover arcs have unbounded capacity and gain $1$.
	Thus we obtain the \emph{condensed time-expanded network} $\mathcal{N}^T/\Delta$,
	whose terminals are $s_0$ and $d_{K-1}$ and whose rounded horizon
	is $K\Delta$. The network $\NT$ is the case $\Delta=1$.
	Figure~\ref{pr:time-condensation} sketches a dynamic network together
	with its time-expanded and condensed time-expanded networks.
	
	\begin{lemma}\label{pr:grid-correspondence}
		Assume that \(T\) and all arc transit times are integer multiples of \(\Delta\). For every feasible flow over time in \(\mathcal{N}\), there exists a feasible static flow in \(\mathcal{N}^T/\Delta\) with the same total source outflow and total sink inflow, and vice versa.
	\end{lemma}
	\begin{proof}
		Given a feasible flow over time $f$ in $\mathcal{N}$, construct a static flow $x$ in $\mathcal{N}^{T}/\Delta$ by
		\[
		x_{a_{\theta}} := \int_{(\theta-1) \Delta}^{\theta\Delta} f_a(t)\, dt, 
		\quad \forall\, a \in A,\; \theta = 1,\dots, K . \notag
		\]
		Given a feasible static flow $x$ on $\mathcal{N}^{T}/\Delta$, construct a dynamic flow $f$ on $\mathcal{N}$ by
		\[
		f_a(t) := \frac{x_{a_{\lceil t/\Delta \rceil}}}{\Delta}, 
		\quad \forall\, a \in A,\; t \in [0,T). \notag
		\]
		For a more detailed proof, see Gro\ss~\cite[Chapter~4]{gross2014approximation}.
		
	\end{proof}
	
	This paper proceeds based on the following assumption.
	\begin{assumption}\label{ass:no-zero-cycle}
		The network $\mathcal{N}$ contains no zero-length cycles.
	\end{assumption}
	
	This assumption ensures that the time-expanded network is acyclic and can be made without loss of generality: zero-length unit-gain or flow-absorbing cycles neither simulate storage nor increase flow; thus, in our optimization problems, flow particles can avoid traversing such cycles without degrading the solution quality. Moreover, zero-length flow-generating cycles are physically unrealistic, as they create flow without source input.
	
	\section{Complexity}\label{sec:complexity}
	We first establish that the GMFP remains NP-hard under standard binary encoding, even on lossy TTSP networks. We then consider the GMTRFP. This problem is APX-hard on general networks and remains NP-hard to solve exactly on lossy TTSP networks. Throughout this section, every integer is encoded in binary and every rational gain is given explicitly by its numerator and denominator.
	
	\subsection{Complexity of the GMFP}\label{sec:complexity-gmfp}
	Our reduction uses \emph{Exact Cover by 3-Sets} (X3C)~\cite{garey1979computers}. An instance consists of a set $U=\{1,\ldots,3k\}$ and a family $\mathcal S=\{S_1,\ldots,S_N\}$ of three-element subsets of $U$. The question is whether some subfamily covers every element of $U$ exactly once.
	
	We will encode set selection by products and use integer transit times to separate products below and above a prescribed threshold. The following elementary approximation lemma supplies these transit times.
	
	\begin{lemma}\label{cx:logarithm}
		Given integers $A\ge2$ and $C\ge1$, an integer $L$ satisfying $|L-C\ln A|\le3/4$ can be computed in time polynomial in $\log A+\log C$.
	\end{lemma}
	\begin{proof}
		
		Write $A=2^h y$, where $h=\lfloor\log_2 A\rfloor$ and $1\le y<2$,
		and set $z=(y-1)/(y+1)$. Then $0\le z<1/3$ and
		$\ln A=h\ln2+\ln y$. The Taylor expansions of $\ln(1+z)$ and
		$\ln(1-z)$ give $\ln y=2\sum_{j=0}^{\infty}z^{2j+1}/(2j+1)$.
		For every integer $J\ge0$, the truncation error satisfies
		\[
		0\le \ln y-2\sum_{j=0}^{J}\frac{z^{2j+1}}{2j+1}
		\le 2\sum_{j=J+1}^{\infty}z^{2j+1}
		=\frac{2z^{2J+3}}{1-z^2}
		\le\frac{1}{12\cdot9^J}.
		\]
		The same expansion and error bound apply to $\ln2$ with $z=1/3$.
		
		Choose the smallest nonnegative integer $J$ such that
		$3\cdot9^J\ge C(h+1)$, and define
		\[
		r=2\sum_{j=0}^{J}
		\frac{h\,3^{-(2j+1)}+z^{2j+1}}{2j+1}.
		\]
		The preceding bounds yield
		$0\le\ln A-r\le(h+1)/(12\cdot9^J)\le1/(4C)$.
		
		It remains to verify the bit complexity. The integer $h$ is obtained
		from the binary length of $A$, and
		$J=O(\log C+\log(h+1))$ can be found by repeated multiplication by $9$.
		Moreover, $z=(A-2^h)/(A+2^h)$ has numerator and denominator of
		$O(\log A)$ bits. Each of the $J+1$ rational terms defining $r$ therefore
		has numerator and denominator of
		$O((J+1)\log A+\log(J+2))$ bits. Even direct common-denominator
		summation keeps all intermediate bit lengths polynomial in
		$\log A+\log C$.
		
		Thus both $r$ and $L=\lfloor Cr+1/2\rfloor$ are computable in the
		claimed time. Finally,
		$|L-C\ln A|\le |L-Cr|+C|r-\ln A|\le1/2+1/4=3/4$.
		
	\end{proof}
	
	Take distinct primes $q_1,\ldots,q_{3k}$ and set
	\[
	a_i=\prod_{j\in S_i}q_j,\qquad B=\prod_{j=1}^{3k}q_j,
	\qquad Q=\prod_{i=1}^{N}a_i.
	\]
	If an element belongs to no set, the instance is immediately negative; hence assume that every element occurs, so $B\le Q$. Unique factorization gives
	\begin{equation}\label{cx:exact-cover-product}
		\prod_{i\in I}a_i=B
		\quad\Longleftrightarrow\quad
		\{S_i:i\in I\}\text{ is an exact cover of }U.
	\end{equation}
	The first $3k$ primes can be generated in polynomial time, and $a_i$, $B$, and $Q$ have polynomial bit lengths.
	
	Let $C=8(B+1)(N+1)$. Apply Lemma~\ref{cx:logarithm} to obtain integers $\ell_i$ and $L_B$ such that $|\ell_i-C\ln a_i|\le3/4$ and $|L_B-C\ln B|\le3/4$. Define $H=L_B+N+1$ and $T=H+1$. For $I\subseteq\{1,\ldots,N\}$, write $p_I=\prod_{i\in I}a_i$ and $L_I=\sum_{i\in I}\ell_i$.
	
	\begin{lemma}\label{cx:time-threshold}
		For every $I\subseteq\{1,\ldots,N\}$, $L_I\le H$ if and only if $p_I\le B$.
	\end{lemma}
	\begin{proof}
		If $p_I\le B$, then
		\[
		L_I\le C\ln B+\frac{3N}{4}
		\le L_B+\frac{3(N+1)}4<H.
		\]
		If $p_I\ge B+1$, then $\ln(1+x)\ge x/(1+x)$ for $x\ge0$ yields
		\[
		L_I-H\ge C\ln\left(1+\frac1B\right)-\frac{7(N+1)}4
		\ge\frac{C}{B+1}-\frac{7(N+1)}4
		=\frac{25(N+1)}4>0.
		\]
		Since $p_I$ is an integer, these two cases exhaust all possibilities.
	\end{proof}
	
\begin{figure}[t]
		\centering
		\includegraphics[width=\linewidth]{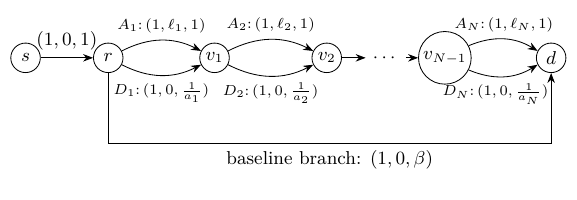}
		\caption{The reduction for the GMFP. The selection modules form a series composition, placed in parallel with the baseline branch. Removing the baseline branch gives the network used for the lossy GMTRFP reduction. $\beta=\frac{2B-1}{2Q}$ and arc labels give $(u,\tau,\gamma)$.}
		\label{fig:gmfp-reduction}
	\end{figure}

	Construct a network with a common source arc $(s,r)$ of capacity $1$, transit time $0$, and gain $1$. Between $r$ and $d$, place two branches in parallel. The first consists of a baseline arc with capacity $1$, transit time $0$, and gain $\beta=(2B-1)/(2Q)$. The second consists of $N$ selection modules in series. Module $i$ contains two parallel arcs $A_i$ and $D_i$ with attributes
	$
	(u,\tau,\gamma)_{A_i}=(1,\ell_i,1) \text{ and }
	(u,\tau,\gamma)_{D_i}=(1,0,1/a_i),
	$ respectively.
	Figure~\ref{fig:gmfp-reduction} illustrates the construction. Parallel arcs can be subdivided by auxiliary nodes, with the added arcs having capacity $1$, transit time $0$, and gain $1$. Thus the construction also satisfies the graph conventions in Section~\ref{sec:preliminaries}.
	
	A selection path $P_I$ uses $A_i$ precisely when $i\in I$. Its transit time is $L_I$, and its gain is $p_I/Q$. All gains are at most $1$, all capacities are $1$, and the network is an acyclic TTSP graph. Its encoding length is polynomial: $\log C=O(\log B+\log N)$, $\log Q=\sum_i\log a_i$, and Lemma~\ref{cx:logarithm} computes all transit times with polynomially many bits.
	
	\begin{theorem}\label{cx:gmfp-hard}
		The GMFP is NP-hard under standard binary encoding, even on acyclic lossy TTSP networks with unit capacities.
	\end{theorem}
	\begin{proof}
		Suppose first that the X3C instance is negative. Consider a feasible flow in the time-expanded network. Since this network is acyclic, its generalized path decomposition consists of source--sink paths. Each component uses either the baseline branch or a selection path $P_I$, possibly together with holdover arcs of gain $1$. A component using $P_I$ and arriving before $T$ must satisfy $L_I\le T-1=H$. Lemma~\ref{cx:time-threshold} implies $p_I\le B$. By~\eqref{cx:exact-cover-product}, equality is impossible; hence $p_I\le B-1$. Its gain is consequently at most $(B-1)/Q<\beta$. A baseline component has gain exactly $\beta$. Summing over the decomposition, the sink inflow is at most $\beta$ times the total flow entering the common source arc. This arc admits at most $T$ units in the whole horizon. Sending at rate $1$ through the baseline branch throughout $[0,T)$ attains the bound, so the optimum is $T\beta$.
		
		Now suppose that $I$ encodes an exact cover. Then $p_I=B$ and $L_I\le H=T-1$. Send flow at source rate $1$ along $P_I$ during $[0,1)$, and along the baseline branch during $[1,T)$. All prefix gains on $P_I$ are at most $1$, so every unit-capacity constraint is satisfied. The final arrival interval of $P_I$ is contained in $[0,T)$. The resulting storage-free flow has value
		\[
		\frac{B}{Q}+(T-1)\beta=T\beta+\frac1{2Q}.
		\]
		Thus the optimum is below $K=T\beta+1/(4Q)$ for negative instances and above $K$ for positive instances, completing the reduction.
	\end{proof}
	
	\subsection{Complexity of the GMTRFP}\label{sec:complexity-gmtrfp}
	We first give an approximation-preserving reduction from the minimum traveling salesman problem with distances in $\{1,2\}$, which is APX-hard~\cite{papadimitriou1993traveling}.
	
\begin{figure}[t]
		\centering
		\includegraphics[width=.9\linewidth]{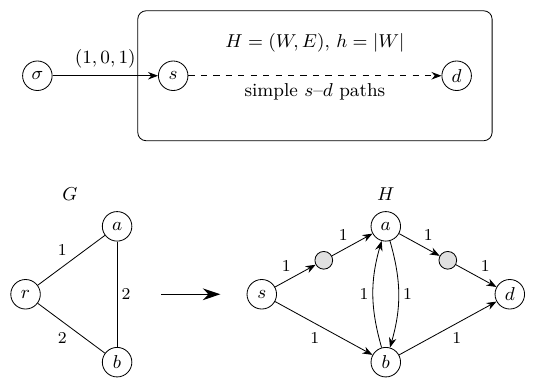}
		\caption{Top: the GMTRFP instance with $T=2h$; labels give
			$(u,\tau,\gamma)$, and every arc of $H$ has attributes $(2^h,1,2)$.
			Bottom: a three-city instance with distinguished city $r$ and the
			resulting $H$; labels give original edge costs and unit arc lengths,
			respectively.}
		\label{fig:gmtrfp-apx-reduction}
	\end{figure}

	\begin{theorem}\label{cx:gmtrfp-apx}
		The GMTRFP is APX-hard under standard binary encoding.
	\end{theorem}
	\begin{proof}
		Let a complete undirected graph on $q\ge3$ cities be a traveling salesman instance, with edge costs $c_{uv}\in\{1,2\}$ and minimum tour cost $C^*$. Replace each edge $uv$ by two oppositely directed arcs, each of weight $3-c_{uv}\in\{1,2\}$. Split a distinguished city into a source $s$ and a sink $d$: its outgoing arcs now leave $s$, and its incoming arcs now enter $d$. Replace every weight-$2$ arc by a two-arc directed path through an auxiliary node, and keep every weight-$1$ arc unchanged. Let $H$ denote the resulting unweighted directed graph with $h$ nodes. A schematic of $H$ is shown in Fig.~\ref{fig:gmtrfp-apx-reduction}. 
		
		A simple $s$--$d$ path in $H$ traverses each auxiliary node together with its unique incoming and outgoing arcs. Contracting these pairs gives a simple partial tour through the distinguished city, with weight equal to the number of arcs in the path. A missing city can be inserted between two consecutive cities without decreasing the weight: the replaced arc has weight at most $2$, whereas the two new arcs have total weight at least $2$. Repeating this operation completes the partial tour to a Hamiltonian tour. Conversely, every Hamiltonian tour of cost $C$ gives a simple $s$--$d$ path of length $3q-C$. It follows that the maximum simple $s$--$d$ path length in $H$ is $j^*=3q-C^*$.
		
		Add a new source $\sigma$ and an arc $(\sigma,s)$ with $(u,\tau,\gamma)=(1,0,1)$. Give every arc of $H$ capacity $2^h$, transit time $1$, and gain $2$, and take horizon $T=2h$; see Figure~\ref{fig:gmtrfp-apx-reduction}. Every simple $\sigma$--$d$ path consists of the common source arc followed by $j\le h-1$ arcs of $H$. Its total transit time is $j$ and its total gain is $2^j$, so its unit-rate temporally repeated value is $F(j)=(2h-j)2^j$. For $0\le r\le h-2$,
		\[
		\frac{F(r+1)}{F(r)}
		=\frac{2(2h-r-1)}{2h-r}
		\ge\frac32.
		\]
		Thus $F$ is strictly increasing.
		
		The source capacity imposes $\sum_Px_P\le1$. Since the prefix gain at any arc of $H$ is at most $2^{h-2}$, the total rate entering such an arc is at most $2^{h-2}\sum_Px_P\le2^{h-2}<2^h$. Hence all other capacity constraints follow from this inequality. Writing $j_P$ for the number of arcs of $H$ on path $P$, every feasible temporally repeated flow therefore has value
		\[
		\sum_Px_PF(j_P)
		\le F(j^*)\sum_Px_P
		\le F(j^*).
		\]
		Conversely, sending at unit rate along a path with $j^*$ arcs of $H$ is feasible and attains $F(j^*)$. Thus the GMTRFP optimal value is $F(j^*)$.
		
		For any $\rho\ge1$, given a feasible temporally repeated flow of value at least $F(j^*)/\rho$, choose a longest path among those with positive path rate, and let $j$ be its length in $H$. By the monotonicity of $F$ and the source constraint, the value of this flow is at most $F(j)$. The ratio bound above therefore gives
		\[
		\left(\frac32\right)^{j^*-j}
		\le\frac{F(j^*)}{F(j)}
		\le\rho,
		\qquad
		j^*-j
		\le\frac{\ln\rho}{\ln(3/2)}
		\le3(\rho-1).
		\]
		The last inequality follows from $\ln\rho\le\rho-1$ and $\ln(3/2)\ge1/3$. Complete the decoded partial tour by the insertion procedure above. Its cost $C$ satisfies $C\le3q-j$, and hence
		\[
		\frac{C}{C^*}
		\le\frac{3q-j^*+j^*-j}{C^*}
		\le1+3(\rho-1).
		\]
		The instance transformation and the solution transformation therefore form an AP-reduction with constant $3$. Both are polynomial-time procedures: $H$ has $O(q^2)$ vertices, and each capacity $2^h$ has $h+1$ bits. This proves APX-hardness.
	\end{proof}
	
	For lossy networks, we reduce from \emph{Product Partition}~\cite{ng2010product}.
	Given integers $a_1,\ldots,a_N$ with $a_i\ge2$, this problem asks whether
	some $I\subseteq\{1,\ldots,N\}$ satisfies
	\[
	\prod_{i\in I}a_i=\prod_{i\notin I}a_i.
	\] Put $Q=\prod_{i=1}^{N} a_i$. If $Q$ is not a perfect square, the instance is negative. We may therefore assume that $Q$ is a perfect square; its integer square root can be computed in polynomial time.
	
	Set $E_0=N+1$ and $C=1024E_0Q^2$. Use Lemma~\ref{cx:logarithm} to compute $\ell_i,L_Q$ satisfying $|\ell_i-C\ln a_i|\le3/4$ and $|L_Q-C\ln Q|\le3/4$, and set $T=C+\lfloor L_Q/2\rfloor$. The reduction uses the network in Figure~\ref{fig:gmfp-reduction} with the baseline branch removed.
	
	\begin{theorem}\label{cx:gmtrfp-lossy-hard}
		The GMTRFP is NP-hard to solve exactly under standard binary encoding, even on acyclic lossy TTSP networks with unit capacities.
	\end{theorem}
	\begin{proof}
		A path $P_I$ chooses the transit-time alternative in module $i$ precisely when $i\in I$. Define $p_I=\prod_{i\in I}a_i$ and $z_I=\ln(p_I/\sqrt Q)$. Its gain is $p_I/Q=e^{z_I}/\sqrt Q$, and its transit time is $\sum_{i\in I}\ell_i$. Since $L_Q$ is an integer,
		\[
		\begin{aligned}
			\left|\left\lfloor\frac{L_Q}{2}\right\rfloor-\frac C2\ln Q\right|
			\le
			\left|\left\lfloor\frac{L_Q}{2}\right\rfloor-\frac{L_Q}{2}\right|
			+\frac12|L_Q-C\ln Q|
			\le\frac12+\frac38=\frac78.
		\end{aligned}
		\]
		Set $\eta_I=\lfloor L_Q/2\rfloor-(C/2)\ln Q-\sum_{i\in I}(\ell_i-C\ln a_i)$. Then $|\eta_I|\le7/8+3|I|/4\le7/8+3N/4<E_0$. Substituting the definitions of $T$ and $z_I$ gives $T-\tau(P_I)=C(1-z_I)+\eta_I$. The value of this path at unit source rate is $(V_I)_+$, where
		\begin{equation}\label{cx:product-value}
			\begin{aligned}
				V_I
				=(T-\tau(P_I))\frac{p_I}{Q}
				=\frac{[C(1-z_I)+\eta_I]e^{z_I}}{\sqrt Q}
				=\frac{C\phi(z_I)+\eta_I e^{z_I}}{\sqrt Q},
			\end{aligned}
		\end{equation}
		and $\phi(z)=(1-z)e^z$.
		
		All paths share the unit-capacity source arc, and every other prefix gain is at most $1$. Hence all remaining capacity constraints follow from $\sum_Px_P\le1$. The objective is $\sum_Ix_{P_I}(V_I)_+$, whose maximum is $\max_I(V_I)_+$.
		
		The function $\phi$ is increasing on $(-\infty,0]$ and decreasing on $[0,\infty)$, because $\phi'(z)=-ze^z$. To bound $|z_I|$ when $p_I^2\ne Q$, w.l.o.g., we may assume $p_I^2>Q$. Integrality then gives
		\[
		|z_I|
		=\frac12\ln\frac{p_I^2}{Q}
		\ge\frac12\ln\left(1+\frac1Q\right)
		\ge\frac1{2(Q+1)}
		\ge\frac1{4Q},
		\]
		where we use $\ln(1+x)\ge x/(1+x)$ for $x\ge0$.
		
		Let $\delta=1/(64Q^2)$. For $0\le z\le1/2$, we have $1-\phi(z)=\int_0^z te^t\,dt\ge z^2/2$. For $-1/2\le z\le0$, we have $1-\phi(z)=\int_z^0(-t)e^t\,dt\ge z^2/4$, since $e^t\ge e^{-1/2}>1/2$ on the integration interval. These estimates and monotonicity outside $[-1/2,1/2]$ imply $\phi(z_I)\le1-\delta$ whenever $p_I^2\ne Q$, for either sign of $z_I$.
		
		Moreover, $V_I>0$ implies $C(1-z_I)+\eta_I>0$, so $z_I<1+\eta_I/C<1+E_0/C<2$ and hence $e^{z_I}<8$. Since $C\delta=16E_0$, Equation~\eqref{cx:product-value} yields
		\[
		V_I
		\le\frac{C(1-\delta)+8E_0}{\sqrt Q}
		=\frac{C-8E_0}{\sqrt Q}
		\qquad\text{if }p_I^2\ne Q\text{ and }V_I>0.
		\]
		It follows that if the \emph{Product Partition} instance is negative, every path is unbalanced, so the optimum is at most $(C-8E_0)/\sqrt Q$. If the  instance is positive, there is a balanced path with $z_I=0$ and $V_I\ge(C-E_0)/\sqrt Q>0$. Thus the rational threshold $K=(C-4E_0)/\sqrt Q$ separates positive and negative \emph{Product Partition} instances. Since $\sqrt Q$ is an integer, the threshold has polynomial bit length, as do $C$, $T$, and the gains $1/a_i$. An exact GMTRFP value algorithm would therefore decide \emph{Product Partition} in polynomial time.
	\end{proof}

	\section{Storage-Free Structure of the OGFP}\label{sec:structure}
	We now establish that the OGFP admits an optimal solution without
	intermediate storage. In lossy networks, such a solution can additionally
	be chosen to use only simple source--sink paths. We first show why the
	temporally repeated construction for classical flows over time does not
	extend to generalized flows.
	
\begin{figure}[t]
		\centering
		\includegraphics[width=.57\linewidth]{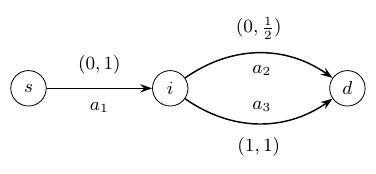}
		\caption{The network in Example~\ref{st:example}. Arc labels give
			$(\tau_a,\gamma_a)$; all capacities are $1$.}
		\label{st:fig-example}
	\end{figure}

	\begin{example}\label{st:example}
		Consider the TTSP network with vertices $s,i,d$ and arcs
		$a_1=(s,i)$ and $a_2,a_3=(i,d)$; see Fig.~\ref{st:fig-example}. All capacities are $1$, the horizon is
		$T=2$, and
		\[
		(\tau_{a_1},\gamma_{a_1})=(0,1),\qquad
		(\tau_{a_2},\gamma_{a_2})=(0,\tfrac12),\qquad
		(\tau_{a_3},\gamma_{a_3})=(1,1).
		\]
		Send flow at rate $1$ along $a_1a_3$ during $[0,1)$ and along
		$a_1a_2$ during $[1,2)$. This  flow delivers $3/2$ units.
		A temporally repeated flow has source rates $x_1,x_2$ on the paths
		$a_1a_3,a_1a_2$, respectively. The static capacity constraint is
		$x_1+x_2\le1$, and its delivered amount is
		$(2-1)x_1+\tfrac12(2-0)x_2=x_1+x_2\le1$.
		Consequently, no temporally repeated flow is optimal.
	\end{example}
	
	\subsection{General networks}\label{st:general}
	Consider the time-expanded formulation $\OGFPT$ of the OGFP.
	Let $X$ be its feasible set and $X^*$ its optimal solution set.
	The set $X$ is compact. Movement flows are bounded by their capacities;
	intermediate and sink holdover flows are bounded by total movement
	inflow, and source holdover flows are bounded by total source movement
	outflow. Feasibility therefore implies that $X^*$ is a nonempty compact
	convex set.
	
	Recall that Assumption~\ref{ass:no-zero-cycle} makes $\NT$ acyclic. It follows that every static flow in $\NT$ can be decomposed into generalized $s_0$--$d_{T-1}$ path flows.

	\begin{theorem}\label{st:storage}
		There exists an optimal solution to the \mbox{OGFP} without intermediate storage.
	\end{theorem}
	
	\noindent\emph{Proof.}
	It suffices to find an optimal solution to $\OGFPT$ with no intermediate
	holdover flow. For any flow $x$, define the “net inflow” through movement arcs and an auxiliary function as follows:
	\[
	\gx(i_\theta)=
	\sum_{a_\eta\in\delta^-(i_\theta)\cap\MT}\gamma_a x_{a_\eta}
	-\sum_{a_\theta\in\delta^+(i_\theta)\cap\MT}x_{a_\theta},
	\qquad
	F(x)=\sum_{i\in V\setminus\{s,d\}}\sum_{\theta\in\TT}
	|\gx(i_\theta)|.
	\]
	Clearly, $F(x)=0$ if and only if no intermediate
	holdover arc is used.
	
	It is easy to verify that $F(x)$ is convex; therefore $F(x)$ attains a minimum on $X^*$ at some $x^0$.
	Suppose $F(x^0)>0$. Let $\theta_1$ be the smallest time with
	$\widehat{x^0}(l_{\theta_1})>0$ at some intermediate node $l$,
	and let $\theta_2$ be the first time with
	$\widehat{x^0}(l_{\theta_2})<0$.
	Nonnegative inventories and zero final inventory imply
	$\theta_2>\theta_1$, and every holdover arc from $l_{\theta_1}$ to
	$l_{\theta_2}$ carries positive flow. Set $r=\theta_2-\theta_1$.
	
	Among the flows in $X^*$ with the same movement net inflows as $x^0$
	at every node of $\NT$, choose $x^*$ minimizing
	\[
	H(x)=\sum_{a_\theta\in\MT}\rho^\theta
	|x_{a_\theta}-x_{a_{\theta-r}}|,
	\qquad 0<\rho<1.
	\]
	This minimum exists by compactness. Copies of movement arcs outside
	their defined time range are assigned flow zero. Write
	$d_{a,\theta}=x^*_{a_\theta}-x^*_{a_{\theta-r}}$.
	Call $a_\theta$ an \emph{excess arc} if $d_{a,\theta}>0$ and a
	\emph{deficit arc} if $d_{a,\theta}<0$. 
	
	Define the \emph{comparison graph} $D$ by taking every excess arc in
	its forward direction and every deficit arc in its reverse direction.
	Zero differences contribute no arc. The arcs of $D$ are called
	\emph{comparison arcs}. An excess arc has comparison
	flow $d_{a,\theta}$ and gain $\gamma_a$; a reversed deficit arc has
	comparison flow $-\gamma_a d_{a,\theta}$ and gain $1/\gamma_a$.
	Let $S$ consist of the nodes reachable from $l_{\theta_2}$ in $D$,
	together with all comparison arcs between them. We establish three claims.
	
	\begin{proofclaim}\label{st:balance}
		Every node of $S$ lies on a time layer at least $r$.
		The node $l_{\theta_2}$ has an outgoing comparison arc.
		If $k_\theta\in S$ has no outgoing comparison arc, then
		\begin{equation}\label{st:sinkgap}
			\widehat{x^*}(k_\theta)-\widehat{x^*}(k_{\theta-r})>0.
		\end{equation}
	\end{proofclaim}
	\begin{proof}
		There is no deficit arc departing before time $r$, since its earlier
		copy has flow zero. Forward excess arcs do not decrease time, and
		backward deficit arcs cannot lead below layer $r$.
		Since $\theta_2\ge r$, this proves the first assertion.
		
		For $i_\theta$ with $\theta\ge r$, direct subtraction gives
		\begin{equation}\label{st:comparison-balance}
			\begin{aligned}
				\widehat{x^*}(i_\theta)-\widehat{x^*}(i_{\theta-r})
				&=
				\sum_{a_\xi\in\delta^-(i_\theta)\cap\MT}
				\gamma_a d_{a,\xi}
				-\sum_{a_\theta\in\delta^+(i_\theta)\cap\MT}
				d_{a,\theta}
				+
				\sum_{\substack{a\in\delta^+(i)\\\theta+\tau_a\ge T}}
				x^*_{a_{\theta-r}}.
			\end{aligned}
		\end{equation}
		The first two sums give the weighted net inflow of the comparison
		flow at $i_\theta$. The last sum is nonnegative and accounts for
		earlier outgoing arc copies whose later copies lie beyond the
		time horizon.
		
		At $l_{\theta_2}$, the left-hand side is negative, so an outgoing
		comparison arc exists. If $k_\theta\in S$ has no outgoing comparison
		arc, then $k_\theta\ne l_{\theta_2}$, and a path from $l_{\theta_2}$
		provides a positive incoming comparison arc. Its net
		inflow is therefore positive, and \eqref{st:sinkgap} follows
		from \eqref{st:comparison-balance}.
	\end{proof}
	
\begin{figure}[t]
		\centering
		\includegraphics[width=.94\linewidth]{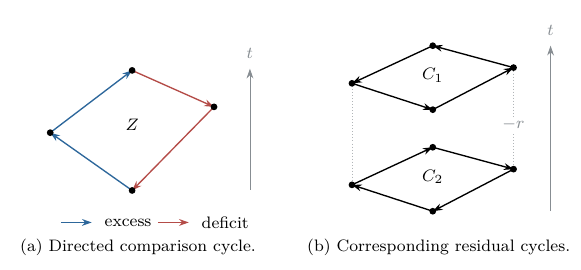}
		\caption{A directed cycle $Z$ in the comparison graph gives a residual
			cycle $C_1$ with the opposite orientation and a translated residual cycle
			$C_2$ with the same orientation. Their gains satisfy
			$\gamma(C_1)\gamma(C_2)=1$.}
		\label{st:comparison-cycle}
	\end{figure}

	\begin{proofclaim}\label{st:acyclic}
		The directed graph $S$ contains no directed cycle.
	\end{proofclaim}
	\begin{proof}
		Suppose $Z$ is a directed cycle in $S$. Since $\NT$ is acyclic,
		$Z$ contains both forward excess arcs and reverse deficit arcs.
		For each movement arc $a_\theta$ involved in $Z$, construct residual
		arcs as follows:
		\[
		\begin{array}{c|cc}
			& C_1 & C_2\\\hline
			d_{a,\theta}>0 & \overleftarrow{a_\theta} & a_{\theta-r}\\
			d_{a,\theta}<0 & a_\theta & \overleftarrow{a_{\theta-r}}.
		\end{array}
		\]
		These arcs form a directed cycle $C_1$ in the opposite direction to
		$Z$ and a directed cycle $C_2$ in the direction of its translate by
		$-r$; see Fig.~\ref{st:comparison-cycle}.
		All have positive residual capacity, and
		$\gamma(C_1)\gamma(C_2)=1$.
		
		If $\gamma(C_1)>1$, choose an excess arc of $Z$. Its original flow
		is positive, so reversing a positive-flow path from $s_0$ to its tail
		gives a path of positive residual capacity from $C_1$ to $s_0$.
		Augmenting around $C_1$ and withdrawing flow along this path reduces
		the source supply, contradicting optimality.
		If $\gamma(C_1)<1$, then $\gamma(C_2)>1$. A reversed deficit arc
		in $Z$ similarly yields a path of positive residual capacity from
		$C_2$ to $s_0$, along which flow can be withdrawn to reduce the
		source supply, again contradicting optimality.
		
		It remains to consider $\gamma(C_1)=1$. Augment around $C_1$ by a
		nonzero circulation $z$, and let $z_{a,\theta}$ denote
		the resulting change in $x^*_{a_\theta}$. Set $z_{a,\theta}=0$ on
		unchanged or nonexistent copies. This augmentation withdraws flow from excess arcs and adds flow to
		deficit arcs, so each nonzero $z_{a,\theta}$ has the opposite sign
		to $d_{a,\theta}$.
		Choose the augmentation small enough to respect all residual
		capacities and satisfy $|z_{a,\theta}|\le|d_{a,\theta}|$. Thus
		$|d_{a,\theta}+z_{a,\theta}|
		=|d_{a,\theta}|-|z_{a,\theta}|$.
		Since the cycle has gain $1$, every movement net inflow is preserved,
		so the resulting flow remains in the set over which $H$ is minimized.
		
		Consequently, we have
		\begin{align*}
			H(x^*+z)-H(x^*)
			&=
			\sum_{a_\theta\in\MT}\rho^\theta
			\left(
			|d_{a,\theta}+z_{a,\theta}-z_{a,\theta-r}|
			-|d_{a,\theta}|
			\right)\\
			&\le
			\sum_{a_\theta\in\MT}\rho^\theta
			\left(
			|d_{a,\theta}+z_{a,\theta}|
			+|z_{a,\theta-r}|-|d_{a,\theta}|
			\right)\\
			&=
			-\sum_{a_\theta\in\MT}\rho^\theta|z_{a,\theta}|
			+\sum_{a_\theta\in\MT}\rho^\theta|z_{a,\theta-r}|\\
			&=
			-\sum_{a_\theta\in\MT}\rho^\theta|z_{a,\theta}|
			+\rho^r
			\sum_{\substack{a_\theta\in\MT\\a_{\theta+r}\in\MT}}
			\rho^\theta|z_{a,\theta}|\\
			&\le
			-(1-\rho^r)
			\sum_{a_\theta\in\MT}\rho^\theta|z_{a,\theta}|
			<0.
		\end{align*}
		The first inequality follows from the triangle inequality and $|d_{a,\theta}+z_{a,\theta}|
		=|d_{a,\theta}|-|z_{a,\theta}|$. This contradicts the minimality of $H(x^*)$.
	\end{proof}
	
\begin{figure}[t]
		\centering
		\includegraphics[width=.94\linewidth]{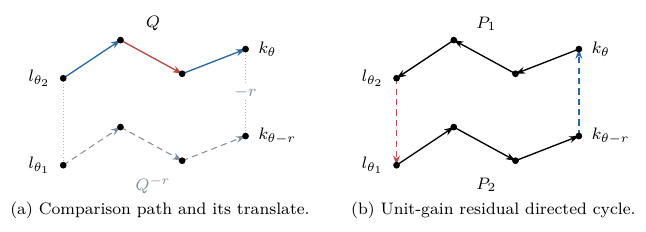}
		\caption{The path exchange. On $Q$, blue arcs are forward excess arcs
			and red arcs are backward deficit arcs. The residual path reverses $Q$
			at the original layers and follows its translate at the lower layers.}
		\label{st:path-exchange}
	\end{figure}

	\begin{proofclaim}\label{st:exchange}
		There is a flow $x\in X^*$ with $F(x)<F(x^*)$.
	\end{proofclaim}
	\begin{proof}
		By Claims~\ref{st:balance} and~\ref{st:acyclic}, a directed path $Q$
		in $S$ leads from $l_{\theta_2}$ to a node $k_\theta$ with no
		outgoing comparison arc. Reverse $Q$ at its original layers to obtain
		a residual path $P_1$ from $k_\theta$ to $l_{\theta_2}$; follow its
		translate by $-r$ to obtain a residual path $P_2$ from $l_{\theta_1}$
		to $k_{\theta-r}$. Both paths lie in the residual network $\mathcal{N}^T_{x^*}$ and
		have positive residual capacities by the excess and deficit
		inequalities. Their gains satisfy $\gamma(P_1)\gamma(P_2)=1$.
		
		Connect $l_{\theta_2}$ to $l_{\theta_1}$ by reverse holdover arcs,
		which have positive residual capacity, and connect $k_{\theta-r}$ to
		$k_\theta$ by forward holdover arcs. This gives a unit-gain directed
		cycle; see Fig.~\ref{st:path-exchange}.
		
		Hence we can place a nonzero circulation $x$ on this cycle. Let $\alpha$ be the amount of flow that $x$ sends out of $k_{\theta}$; then the amount entering $l_{\theta_2}$ and the amount leaving $l_{\theta_1}$ are both $\gamma(P_1)\alpha$, and the amount entering $k_{\theta-r}$ is $\gamma(P_1)\gamma(P_2)\alpha=\alpha$. We restrict
		$
		0<\alpha \le \min\!\left\{-\frac{\widehat{x^*}(l_{\theta_2})}{\gamma(P_1)},\ \frac{\widehat{x^*}(l_{\theta_1})}{\gamma(P_1)},\ \widehat{x^*}(k_{\theta})-\widehat{x^*}(k_{\theta-r})\right\}.
		$
		Clearly, $x^*+x$ is a feasible flow on $\mathcal{N}^T$ and remains an optimal solution to $\mathrm{OGFP}^T$. We have
		\[
		\begin{aligned}
			F(x^*+x)-F(x^*)
			&= \bigl|\widehat{x^*+x}(l_{\theta_2})\bigr|
			+ \bigl|\widehat{x^*+x}(l_{\theta_1})\bigr|
			+ \bigl|\widehat{x^*+x}(k_{\theta})\bigr|
			+ \bigl|\widehat{x^*+x}(k_{\theta-r})\bigr| \\
			&\quad - \Bigl(\bigl|\widehat{x^*}(l_{\theta_2})\bigr|
			+ \bigl|\widehat{x^*}(l_{\theta_1})\bigr|
			+ \bigl|\widehat{x^*}(k_{\theta})\bigr|
			+ \bigl|\widehat{x^*}(k_{\theta-r})\bigr|\Bigr) \\
			&= -2\,\gamma(P_1)\alpha
			+ \bigl|\widehat{x^*}(k_{\theta})-\alpha\bigr|
			+ \bigl|\widehat{x^*}(k_{\theta-r})+\alpha\bigr| \\
			&\quad- \bigl(\,|\widehat{x^*}(k_{\theta})|+|\widehat{x^*}(k_{\theta-r})|\,\bigr) \\
			&\le -2\,\gamma(P_1)\alpha \;<\; 0.
		\end{aligned}
		\]
		The penultimate inequality follows from the fact that for any $\alpha, \beta, \mu \in \mathbb{R}$ with $\beta-\mu \ge \alpha > 0$, we have $|\beta|+|\mu| \ge |\beta-\alpha| + |\mu+\alpha|$.
		
	\end{proof}
	
	Claim~\ref{st:exchange} contradicts the choice of the minimum of $F$
	on $X^*$. Thus $F(x^*)=F(x^0)=0$, and an optimal solution uses no intermediate
	storage. Then we prove the theorem.\qed
	
	We further obtain the following stronger result.
	
	\begin{corollary}\label{st:selection}
		The minimum of $\sum_{a\in A}\int_0^T f_a(t)\,dt$ over all optimal
		OGFP solutions is attained by a flow without intermediate storage.
	\end{corollary}
	\begin{proof}
		The time-expanded correspondence preserves the additional objective,
		which becomes $\sum_{a_\theta\in\MT}x_{a_\theta}$. Repeat the proof
		of Theorem~\ref{st:storage}, restricting $X^*$ to the optimal OGFP$^T$ flows
		minimizing this sum. Claim~\ref{st:balance} is unchanged.
		
		For Claim~\ref{st:acyclic}, if $\gamma(C_1)\ne1$, one of $C_1,C_2$
		has gain greater than $1$, giving the same contradiction to minimum
		source supply. If both gains are $1$, choose sufficiently small
		circulations around $C_1$ and $C_2$ so that their changes in
		$x_{a_\theta}$ and $x_{a_{\theta-r}}$, respectively, have the same
		magnitude and opposite signs. Both circulations preserve source supply, and their
		changes in $\sum_{a_\theta\in\MT}x_{a_\theta}$ sum to zero.
		By minimality, neither change can be negative, so both are zero.
		We then apply only the circulation around $C_1$, which preserves
		this sum and strictly decreases the function $H$ as in the original proof.
		
		For Claim~\ref{st:exchange}, the change in $x_{a_{\theta-r}}$ along
		$P_2$ is the negative of the change in $x_{a_\theta}$ along $P_1$, so their
		contributions to $\sum_{a_\theta\in\MT}x_{a_\theta}$ cancel.
		The resulting flow therefore remains in $X^*$, and the same strict
		decrease of $F$ proves the claim. Consequently, the  argument
		yields a storage-free flow in $X^*$.
	\end{proof}
	
	\subsection{Lossy networks}\label{st:lossy}
	
	\begin{theorem}\label{st:simple}
		In a lossy network, there exists an optimal OGFP solution that is
		storage-free, uses only simple $s$--$d$ paths, and minimizes
		$\sum_{a\in A}\int_0^T f_a(t)\,dt$ among all optimal OGFP solutions.
	\end{theorem}
	
	\begin{proof}
		Let $\PP$ be the set of source--sink paths in $\NT$ that use no
		holdover arc at an intermediate node. By Corollary~\ref{st:selection}
		and the time-expanded correspondence, choose an optimal solution
		$x^*$ to $\OGFPT$ without using intermediate holdover arcs that minimizes
		$\sum_{a_\theta\in\MT}x_{a_\theta}$ among all optimal solutions.
		Decompose it into  path flows $x^*_P$, $P\in\PP$.
		Let $\PP_1\subseteq\PP$ consist of the paths whose projections onto
		$\mathcal{N}$, after omitting source and sink holdovers, are simple
		$s$--$d$ paths. We show that every used path belongs to $\PP_1$.
		
		Suppose $x^*_P>0$ for a path $P\notin\PP_1$. It has a subpath
		$\Gamma$ whose projection is a directed cycle; write
		$P=P_1\cdot\Gamma\cdot P_2$. Replace this path flow by a flow with
		source value $\gamma(\Gamma)x^*_P$ that follows $P_1$, waits at the
		initial node of $\Gamma$ for $\tau(\Gamma)$ time units, and then
		follows $P_2$. Since $\gamma(\Gamma)\le1$, the flow entering each arc of $P_1$
		does not increase. The flow along $\Gamma$ is removed, while each
		arc of $P_2$ receives the same amount of flow at the same time as before. Thus the replacement is feasible in
		$\OGFPT$ and satisfies the same sink demand.
		
		If $\gamma(\Gamma)<1$,  source supply strictly decreases,
		contradicting optimality. If $\gamma(\Gamma)=1$, source supply is
		unchanged, but removing the positive flow along $\Gamma$ strictly
		decreases $\sum_{a_\theta\in\MT}x_{a_\theta}$, contradicting its
		minimality. Hence every used path belongs to $\PP_1$, and the
		corresponding flow over time has all the required properties.
	\end{proof}

	\begin{corollary}\label{st:gmfp}
		The GMFP admits an optimal solution without intermediate storage.
		In a lossy network it admits such a solution using only simple
		$s$--$d$ flow paths.
	\end{corollary}
	\begin{proof}
		Let $b$ be the optimal GMFP value. An optimal OGFP solution for demand
		$b$ delivers the same amount, and Theorems~\ref{st:storage}
		and~\ref{st:simple} give the stated properties.
	\end{proof}

	\section{Storage-Free Approximation in Lossy Networks}\label{sec:approximation}
	We apply the time-condensation algorithm of Fleischer and Skutella~\cite{fleischer2007quickest}, following its extension to lossy generalized flows by Gro\ss{}~\cite[Chapter~5]{gross2014approximation}. Theorem~\ref{st:simple} bounds the rounding error along simple paths and supplies the same path structure for the computed condensed flow, so the usual smoothing and scaling recover a storage-free flow. The algorithm uses the  condensed network from the preliminaries; only intermediate holdover arcs are deleted. Appendix~\ref{app:approximation} gives the details.
	
	\begin{theorem}\label{ap:main}
		For a feasible lossy OGFP instance with demand $b$, horizon $T>0$, and precision $\epsilon>0$, Algorithm~\ref{alg:ogfp} computes a storage-free flow delivering $b$ with source supply at most $S^*(b,T)$ by time $(1+\epsilon)T$. Its running time and output encoding length are polynomial in the input size and $1/\epsilon$.
	\end{theorem}
	
	\begin{algorithm}[t]
		\caption{Storage-free approximation for the OGFP}\label{alg:ogfp}
		\small\renewcommand{\baselinestretch}{1}\selectfont
		\textbf{Input:} A lossy network $\mathcal{N}$ with horizon $T>0$, demand $b\ge0$, and precision $0<\epsilon\le 1$.\par
		\textbf{Output:} A storage-free flow delivering $b$ by $(1+\epsilon)T$, or a certificate that $b>v^*(T)$.
		\begin{enumerate}[leftmargin=*,itemsep=3pt,topsep=4pt]
			\item If $b=0$, return the zero flow. Otherwise, set
			$\delta=\epsilon/10$ and $\Delta=\delta^2T/n$.
			\item Set $\tau'_a=\lceil\tau_a/\Delta\rceil\Delta$ for each arc and
			$T'=\lceil(1+\delta)^3T/\Delta\rceil\Delta$.
			\item Construct the condensed network $\mathcal{N}^{T'}/\Delta$ for the rounded transit times, retaining source and sink holdover arcs and deleting intermediate holdover arcs.
			\item For sink demand $(1+\delta)b$, compute a static flow $x$ minimizing source supply. If infeasible, return $b>v^*(T)$. Fix the minimum source supply and minimize total flow on all movement arcs.
			\item Decompose $x$ into simple path flows and convert them to path-flow rates $f'_P$ for the rounded transit times, as described in Appendix~\ref{app:condensed}.
			\item Route the source rates
			\begin{equation}\label{ap:recovery}
				f_P(t)=\frac{1}{(1+\delta)\delta T'}
				\int_{t-\delta T'}^t f'_P(z)\,dz
			\end{equation}
			along the simple paths, with transit times $\tau$ and no storage, and return $f$.
		\end{enumerate}
	\end{algorithm}
	
	\begin{proof}
		Lemma~\ref{ap:forward} gives feasibility and the source-supply bound.
		Theorem~\ref{st:simple} guarantees simple $s$--$d$ paths for the
		selected static flow. The smoothing, capacity, completion time, and
		encoding estimates are proved in Appendix~\ref{app:approximation}.
	\end{proof}
	
	\begin{corollary}\label{ap:gmfp}
		For every $0<\epsilon<1$, a storage-free generalized flow delivering at
		least $(1-\epsilon)v^*(T)$ by time $(1+\epsilon)T$ can be computed in time
		polynomial in the input size and $1/\epsilon$. 
	\end{corollary}
	\begin{proof}
		
		Find an $s$--$d$ path $P$ of minimum transit time.
		If none exists or $\tau(P)\ge T$, the zero flow is optimal.
		Otherwise, sending flow along $P$ gives the positive feasible value
		$
		b_0=(T-\tau(P))\gamma(P)
		\min_{a\in P}\frac{u_a}{\gamma(a\mid P)}.
		$
		Use Algorithm~\ref{alg:ogfp} as a demand oracle and bisect between
		$b_0$ and $T\sum_{a\in\delta^-(d)}\gamma_a u_a$, retaining the flow
		at the lower endpoint. By Lemma~\ref{ap:forward}, a rejected demand
		exceeds $v^*(T)$, while an accepted demand is delivered without
		storage by time $(1+\epsilon)T$. Stopping when the interval width
		is at most $\epsilon b_0$ therefore gives a retained value of at least
		$
		v^*(T)-\epsilon b_0\ge(1-\epsilon)v^*(T).
		$
		Both bounds have polynomial encoding length, so only polynomially
		many oracle calls are needed. Theorem~\ref{ap:main} gives the claimed
		running time.
		
	\end{proof}

	Given a demand $b>0$ and source budget $B\ge0$, the \emph{quickest generalized flow problem} minimizes the completion time among flows delivering $b$ with supply at most $B$.
	\begin{corollary}\label{ap:quickest}
		The quickest generalized flow problem on lossy networks admits an FPTAS whose output is storage-free.
	\end{corollary}
	\begin{proof}
		Let $T^*(b,B)$ be the infimum feasible completion time.
		Find a maximum-gain $s$--$d$ path $P$.
		If none exists or $B\gamma(P)<b$, the instance is infeasible.
		Otherwise, the bounds
		\[
		T_{\rm low}=\frac{b}{\sum_{a\in\delta^-(d)}\gamma_a u_a},
		\qquad
		T_{\rm up}=\tau(P)+
		\frac{b}{\gamma(P)\min_{a\in P}\{u_a/\gamma(a\mid P)\}}
		\]
		satisfy $0<T_{\rm low}\le T^*(b,B)\le T_{\rm up}$.
		Use Algorithm~\ref{alg:ogfp} with accuracy
		$\eta=\epsilon/3$ and supply bound $(1+\delta)B$
		in both linear programs in step 4 as a horizon oracle.
		By Lemma~\ref{ap:forward}, every rejected horizon is a lower bound
		on $T^*(b,B)$, while an accepted horizon $T$ returns a storage-free
		flow meeting demand $b$ and budget $B$ by time $(1+\eta)T$.
		Bisect $[T_{\rm low},T_{\rm up}]$ until its width is at most
		$\eta T_{\rm low}$, retaining the flow at the accepted upper endpoint.
		Its completion time is at most
		\[
		(1+\eta)\bigl(T^*(b,B)+\eta T_{\rm low}\bigr)
		\le(1+\eta)^2T^*(b,B)
		\le(1+\epsilon)T^*(b,B).
		\]
		Theorem~\ref{ap:main} gives polynomial running time.
		
	\end{proof}
	
	\section{Conclusion}\label{sec:conclusion}
	In this paper, we resolved a fundamental structural question regarding generalized flows over time by proving that intermediate storage is not required for optimality. Furthermore, we showed that optimal flows in lossy networks can use only simple paths. The simple-path property enables storage-free time-approximation algorithms for the OGFP and GMFP, and an FPTAS for the quickest-flow problem with a source-supply budget. We also proved that the GMFP remains NP-hard under standard binary encoding, even on lossy TTSP networks, and that the GMTRFP is APX-hard with general gains and NP-hard on lossy TTSP networks. An interesting direction for future work is to investigate whether similar storage-free or simple-path properties hold for more complex generalized minimum cost flows over time.


	\section*{Funding}
	This work was supported in part by the National Natural Science Foundation of China (Grant Nos. T2341006, 12571382, 12501450, and 2501445) and the National Key R\&D Program of China (Grant No. 2022YFA1003900).
	
	\section*{Declaration of competing interest}
	The authors declare that they
	have no competing interests.
	
	\section*{Data availability}
	No empirical datasets were used in
	this theoretical study.

	\appendix
	\renewcommand{\thesubsection}{\Alph{section}.\arabic{subsection}}
	\renewcommand{\thetheorem}{\Alph{section}.\arabic{theorem}}
	\section{Details of the Approximation Algorithm}\label{app:approximation}
	We give the amplification, smoothing, and scaling estimates from the time-condensation framework of~\cite{fleischer2007quickest,gross2014approximation} for Algorithm~\ref{alg:ogfp}. It suffices to consider $0<\epsilon\le1$. All parameters below are those of the algorithm.
	
	\subsection{The condensed formulation}\label{app:condensed}
	Consider a condensed network $\mathcal{N}^{T'}/\Delta$ without intermediate holdover arcs.  This network has $O(n^2\epsilon^{-2})$ nodes and $O(mn\epsilon^{-2})$ movement arcs, since $T'/\Delta=O(n\epsilon^{-2})$. The source and sink holdover arcs add only $O(n\epsilon^{-2})$ arcs. Assumption~\ref{ass:no-zero-cycle} makes it acyclic.
	
	For demand $(1+\delta)b$, solve two linear programs: first minimize source supply, and then fix that minimum and minimize the sum of movement-arc flows. Using $\Delta$ as the time unit, Corollary~\ref{st:selection} shows that both minima  are attained without intermediate holdovers. Thus the selected optimum decomposes into path flows that project
	onto simple $s$--$d$ paths in the original network.
	
	A path decomposition is obtained by repeatedly subtracting a maximal positive path flow. Each subtraction makes at least one positive arc zero. The number of components is therefore at most the total number of arcs, hence $O(mn\epsilon^{-2})$ when $b>0$. In recording a component, omit its initial source holdovers and final sink holdovers; its departure layer is the layer of its first movement. A component of source amount $x_P$ departing in layer $k$ induces
	a constant source rate $x_P/\Delta$ on $[k\Delta,(k+1)\Delta)$
	along the corresponding $s$--$d$ path in the original network. Combining components corresponding to the
	same path gives source rates $f'_P$ and induced arc rates $f'_a$
	for transit times $\tau'$. These rates respect capacities, since movement capacities are $\Delta u_a$. This is the correspondence of Lemma~\ref{pr:grid-correspondence}, and it preserves source supply and sink inflow.
	
	\subsection{Amplification and smoothing}
	\begin{lemma}\label{ap:forward}
		If $S^*(b,T)<\infty$, the condensed network in Algorithm~\ref{alg:ogfp} admits a flow delivering $(1+\delta)b$ with source supply at most $(1+\delta)S^*(b,T)$.
	\end{lemma}
	\begin{proof}
		For an optimal flow $f$ of the OGFP, replace $T$ and each $\tau_a$ by
		$(1+\delta)^2T$ and $(1+\delta)^2\tau_a$, respectively,
		and use the rate $f_a(t/(1+\delta)^2)$ at time $t$. Supply and delivery are multiplied by $(1+\delta)^2$. With the original transit times, the longer traversals are simulated by waiting at the head of each arc. Thus demand $(1+\delta)^2b$ is feasible by $(1+\delta)^2T$ with supply at most $(1+\delta)^2S^*(b,T)$. By Theorem~\ref{st:simple}, there is a storage-free flow on simple paths meeting this demand with no greater supply. Write its source rates as $f^1_P$ and its induced arc rates as $f^1_a$. 
		
		Each used path is simple and contains at most $n-1$ arcs.
		Since rounding increases each arc transit time by less than $\Delta$,
		for every used path $P$ and every $a\in P$,
		\[
		0\le\tau'(a\mid P)-\tau(a\mid P)\le n\Delta=\delta^2T,
		\qquad
		0\le\tau'(P)-\tau(P)\le n\Delta.
		\]
		Extend the source rates by zero outside their departure intervals
		and define
		\[
		f^2_P(t)=\frac{1}{(1+\delta)\delta T}
		\int_{t-\delta T}^{t}f^1_P(z)\,dz.
		\]
		Send these path flows with transit times $\tau'$ and no waiting.
		For any arc $a$, the resulting input rate satisfies
		\begin{align*}
			f^2_a(t)
			&=\sum_{P\ni a}\gamma(a\mid P)
			f^2_P(t-\tau'(a\mid P))\\
			&=\frac{1}{(1+\delta)\delta T}
			\sum_{P\ni a}\gamma(a\mid P)
			\int_{t-\delta T-\tau'(a\mid P)}^{t-\tau'(a\mid P)}
			f^1_P(z)\,dz\\
			&=\frac{1}{(1+\delta)\delta T}
			\sum_{P\ni a}\gamma(a\mid P)
			\int_{t-\delta T-\tau'(a\mid P)+\tau(a\mid P)}
			^{t-\tau'(a\mid P)+\tau(a\mid P)}
			f^1_P(z-\tau(a\mid P))\,dz\\
			&\le\frac{1}{(1+\delta)\delta T}
			\int_{t-\delta T-n\Delta}^{t}
			\sum_{P\ni a}\gamma(a\mid P)
			f^1_P(z-\tau(a\mid P))\,dz\\
			&=\frac{1}{(1+\delta)\delta T}
			\int_{t-\delta T-n\Delta}^{t}f^1_a(z)\,dz\\
			&\le\frac{\delta T+n\Delta}{(1+\delta)\delta T}u_a
			=u_a.
		\end{align*}
		Moreover, for every used path $P$,
		\[
		\int_{\mathbb R} f^2_P(t)\,dt
		=\frac{1}{(1+\delta)\delta T}
		\int_{\mathbb R}\int_z^{z+\delta T}f^1_P(z)\,dt\,dz
		=\frac{1}{1+\delta}\int_{\mathbb R}f^1_P(z)\,dz.
		\]
		Hence
		\begin{align*}
			\sum_P\int_{\mathbb R}f^2_P(t)\,dt
			&=\frac{1}{1+\delta}
			\sum_P\int_{\mathbb R}f^1_P(t)\,dt
			\le(1+\delta)S^*(b,T),\\
			\sum_P\gamma(P)\int_{\mathbb R}f^2_P(t)\,dt
			&=\frac{1}{1+\delta}
			\sum_P\gamma(P)\int_{\mathbb R}f^1_P(t)\,dt
			=(1+\delta)b.
		\end{align*}
		Since $f^1_P(z)=0$ whenever
		$z+\tau(P)>(1+\delta)^2T$, the smoothing formula gives
		$
		f^2_P(t)=0
		\text{ whenever }
		t>(1+\delta)^2T-\tau(P)+\delta T.
		$
		Thus every positive path flow arrives by
		\begin{align*}
			(1+\delta)^2T+\delta T+\tau'(P)-\tau(P)
			&\le(1+\delta)^2T+\delta T+n\Delta\\
			&=\bigl((1+\delta)^2+\delta+\delta^2\bigr)T\\
			&\le(1+\delta)^3T\le T'.
		\end{align*}
		Integrating over the $\Delta$-intervals therefore gives the required
		static flow by Lemma~\ref{pr:grid-correspondence}, with no
		intermediate holdover flow.
	\end{proof}
	
	\subsection{Recovery and complexity}
	\begin{proof}[Proof of Theorem~\ref{ap:main}]
		Lemma~\ref{ap:forward} gives feasibility and bounds the minimum source
		supply by $(1+\delta)S^*(b,T)$. We fix this minimum when minimizing
		total movement flow. By  Theorem~\ref{st:simple}, the selected optimum
		decomposes into flows along simple $s$--$d$ paths in the original
		network. They respect capacities for $\tau'$
		and deliver $(1+\delta)b$ by $T'$.
		
		Apply~\eqref{ap:recovery} with the original transit times. Since
		$0\le\tau'(a\mid P)-\tau(a\mid P)\le n\Delta$, we obtain
		\begin{align*}
			f_a(t)
			&=\sum_{P\ni a}\gamma(a\mid P)f_P(t-\tau(a\mid P))\\
			&=\frac{1}{(1+\delta)\delta T'}
			\sum_{P\ni a}\gamma(a\mid P)
			\int_{t-\delta T'-\tau(a\mid P)}^{t-\tau(a\mid P)}
			f'_P(z)\,dz\\
			&=\frac{1}{(1+\delta)\delta T'}
			\sum_{P\ni a}\gamma(a\mid P)
			\int_{t-\delta T'-\tau(a\mid P)+\tau'(a\mid P)}
			^{t-\tau(a\mid P)+\tau'(a\mid P)}
			f'_P(z-\tau'(a\mid P))\,dz\\
			&\le\frac{1}{(1+\delta)\delta T'}
			\int_{t-\delta T'}^{t+n\Delta}
			\sum_{P\ni a}\gamma(a\mid P)
			f'_P(z-\tau'(a\mid P))\,dz\\
			&=\frac{1}{(1+\delta)\delta T'}
			\int_{t-\delta T'}^{t+n\Delta}f'_a(z)\,dz\\
			&\le\frac{\delta T'+n\Delta}{(1+\delta)\delta T'}u_a
			\le u_a,
		\end{align*}
		where the last inequality uses $n\Delta=\delta^2T\le\delta^2T'$.
		Moreover, integrating~\eqref{ap:recovery} gives
		$
		\int_{\mathbb R}f_P(t)\,dt
		=\frac{1}{1+\delta}\int_{\mathbb R}f'_P(t)\,dt.
		$
		Consequently,
		\begin{align*}
			\sum_P\int_{\mathbb R}f_P(t)\,dt
			&=\frac{1}{1+\delta}
			\sum_P\int_{\mathbb R}f'_P(t)\,dt
			\le S^*(b,T),\\
			\sum_P\gamma(P)\int_{\mathbb R}f_P(t)\,dt
			&=\frac{1}{1+\delta}
			\sum_P\gamma(P)\int_{\mathbb R}f'_P(t)\,dt
			=b.
		\end{align*}

		Smoothing delays departures by at most $\delta T'$, while restoring the original transit times can only shorten paths. Completion therefore occurs by $(1+\delta)T'$. Moreover,
		\[
		(1+\delta)T'
		\le(1+\delta)^5T
		=\left(1+\frac{\epsilon}{10}\right)^5T
		\le(1+\epsilon)T.
		\]
		
		The condensed network has polynomial size, and its rational coefficients
		have polynomial encoding length. Both linear programs and the path
		decomposition can therefore be computed in polynomial time.
		A path-flow component of value $x_P$ departing in layer $k$ contributes
		\[
		\frac{x_P}{(1+\delta)\Delta\delta T'}
		\bigl(\min\{(k+1)\Delta,t\}
		-\max\{k\Delta,t-\delta T'\}\bigr)_+
		\]
		to $f_P(t)$. This continuous piecewise-linear function has at most
		four breakpoints. Its contribution to $f_a(t)$ is obtained by
		shifting time by $\tau(a\mid P)$ and multiplying by $\gamma(a\mid P)$.
		All component values, breakpoints, $\tau(a\mid P)$, and
		$\gamma(a\mid P)$ have polynomial encoding length. Thus the returned
		flow has polynomial encoding length and can be evaluated in polynomial time.
	\end{proof}

	\end{document}